\documentclass[journal=jctc,manuscript=article]{achemso}
\usepackage{graphicx}
\usepackage{dcolumn}
\usepackage{bm}
\usepackage{gensymb}
\usepackage{braket}
\usepackage[T1]{fontenc} 
\usepackage{amsmath} 
\usepackage{caption}
\usepackage[colorlinks,allcolors=blue,urlcolor=blue]{hyperref}
\DeclareCaptionLabelSeparator{vert}{ $\!$| $\!$}
\usepackage[justification=justified,labelsep=vert,labelfont=bf]{caption}

\newcommand{\autocite}[1]{\cite{#1}}

\makeatletter

\author{Marco Bernardi}
\email{bmarco@caltech.edu}
\affiliation{Department of Applied Physics and Materials Science, California Institute of Technology, Pasadena, CA 91125, USA.}
\affiliation{Department of Physics, California Institute of Technology, Pasadena, CA 91125, USA.}
\title{Efficient Mean-Field Simulation of Quantum Circuits Inspired by Density Functional Theory} 

\begin{document}

\begin{abstract}
\noindent 
Exact simulations of quantum circuits (QCs) are currently limited to $\sim$50 qubits because the memory and computational cost required to store the QC wave function scale exponentially with qubit number. 
Therefore, developing efficient schemes for approximate QC simulations is a current research focus. 
Here we show simulations of QCs with a method inspired by density functional theory (DFT), a widely used approach to study many-electron systems. 
Our calculations can predict marginal single-qubit probabilities (SQPs) with over 90\% accuracy in several classes of QCs with universal gate sets, using memory and computational resources linear in qubit number despite the formal exponential cost of the SQPs. 
This is achieved by developing a mean-field description of QCs and formulating optimal single- and two-qubit gate functionals $-$ analogs of exchange-correlation functionals in DFT $-$ to evolve the SQPs without computing the QC wave function. Current limitations and future extensions of this formalism are discussed.
\end{abstract}

\setcitestyle{super}
%
%
\section{1. Introduction}
Noisy intermediate-scale quantum devices promise exciting advances in quantum algorithms with no classical counterpart~\cite{NISQ, Chuang, Google}. 
Classical simulations remain essential to understand the physics of these quantum devices, improve their design, and accelerate their progress~\cite{juqs,qhipster,quest,64qubit}. An important direction is the development of approximate schemes that are both accurate and computationally efficient, 
enabling simulations of generic QCs with arbitrary depth and degree of entanglement, ideally with favorable computational scaling. 
Work in this area has focused on tensor network matrix product states to simulate QCs with a range of structures, gate types, entanglement, and noise~\cite{Jozsa, Markov2008, Vidal, Yoran, Jozsa-2, PRX, quimb}, and more recently on simulations of generic QCs using neural-network quantum states~\cite{Carleo}. Despite these notable advances, approximate QC simulations remain an area of active investigation.
\\ 
\indent
There is an intriguing parallel between many-electron and many-qubit systems. In the many-electron problem $-$ a grand challenge in chemistry and materials physics~\cite{Reining} $-$ exact solutions are possible only for systems with one electron (the hydrogen atom). Therefore, unlike QC simulations, electronic structure calculations of molecules and materials are dominated by approximate methods~\cite{DFT,CP,QMC,DMRG,DMFT,Reining,Motta2017,Zgid},
among which density functional theory (DFT) is the main workhorse. Leveraging a mean-field description centered on the electron density, DFT achieves low-polynomial scaling with system size, enabling studies of matter with thousands of interacting electrons~\cite{DFT, Burke}.
Methods to study QCs with a similar trade-off of cost and accuracy would be expedient. Early work on relating DFT to QCs focused on formal mapping of QCs onto lattice fermions~\cite{Nori} or connecting time-dependent DFT and spin Hamiltonians~\cite{Guzik}. These notable efforts differ in method and scope from this work.
\\ 
\indent
Here we show a DFT-inspired approach for QCs $-$ in short, QC-DFT $-$ able to accurately simulate single-qubit probabilities (SQPs) in QCs with computational cost scaling linearly with qubit number and depth, despite the formal exponential cost of the SQPs. We present results for various random QCs using two different universal gate sets, and demonstrate the formulation and optimization of QC-DFT gate functionals. We also apply this formalism to nonrandom QCs, studying how the SQP distribution changes with QC size, as well as simulate a simple model Hamiltonian and a quantum algorithm. These results show that even though the exact QC wave function is exponentially complex, marginal probability distributions such as the SQPs can be obtained with a favorable trade-off of cost and accuracy without computing the QC wave function. Although the current formulation is limited to QCs with low entanglement, we discuss an \mbox{extension} based on reduced density matrices which may enable further progress.
\\
\indent
\section{2. Theory: QC-DFT and Gate Functionals}
The QC wave function for $N$ qubits can be expanded in the computational basis as
\begin{equation}
\Psi = \sum_{i_1 i_2 \ldots i_N} c_{i_1 i_2 \ldots i_N} |i_1 i_2 \ldots i_N \rangle = \sum_x^{2^N-1} c_x |x\rangle
\end{equation}
where $i_n = 0,1$ are basis states for a single qubit, $x$ are binary numbers from 0 to $2^{N-1}$, $c_{x}$ are state-vector amplitudes, namely expansion coefficients of the QC wave function, and $|x\rangle=|i_1 i_2 \ldots i_N\rangle$ are $N$-qubit states in the computational basis ($N$-bit long bitstrings). For $N$ qubits, accessing this wave function requires storing and manipulating  $2^N$ complex numbers, which is out of reach for modern computers for $N>50$. (A laptop can handle $N \approx 25$ qubits, and a small computer cluster $N \approx 30$ on a single core; parallelization is needed beyond $N\!=\!30$.) 
In a gate-based QC, the wave function evolves at each cycle (or step) via a unitary transformation, and it can be computed exactly with a classical algorithm by applying single- and two-qubit gates as $2\times 2$ unitary matrices and updating pairs of amplitudes in place~\cite{juqs,qhipster}. 
From the exact wave function at step $s$, one can obtain the $N$-qubit probability distribution 
$\tilde{P}_s(x) = |\langle x |\Psi_s\rangle|^2$, 
which can be measured experimentally but is exponentially hard to compute~\cite{Google}. 
\\
\indent 
Here we take a different approach and focus on the evolution of each individual qubit as a result of mean-field interactions with single- and two-qubit gates. 
We define the \textit{single-qubit probability (SQP)} for qubit $n$, with values between 0 and 1, as the probability of measuring qubit $n$ in the excited state $|1\rangle$ at step $s$, regardless of the state of the other qubits:
\begin{equation}
\label{eq:SQPdef}
p_s^{(n)} = \sum_{\{i_q,\, q\neq n\}} |\langle i_1, i_2, \ldots, i_n\!=\!1, \ldots, i_N|\Psi_s\rangle |^2.
\end{equation}
The \textit{exact} SQPs are marginals of the $N$-qubit probability distribution $\tilde{P}_s(x)$, and are also exponentially hard to compute because they require knowledge of the QC wave function.  
We define the SQP vector at step $s$, $\mathbf{p}_s = (p^{(1)}, p^{(2)}, \ldots, p^{(N)})_s\,$, as the set of SQPs for all qubits in the QC. Note that the SQP vector has $N$ components, and thus it can be stored with memory resources linear in qubit number $N$. Experimentally, the SQPs can be accessed by measuring the state of each single qubit. 
\\
\indent  
We model the evolution of the SQP vector $\mathbf{p}_s$ under the effect of single- and two-qubit gates, using an approximate mean-field approach inspired by DFT. 
In a general QC, single- and two-qubit gates are applied to a set of qubits at each step $s$. As a result, the SQP vector evolves to a new value at step $s+1$:
\begin{equation}
\label{eq:evolution}
  \mathbf{p}_{s+1} = f_{\rm G}(\mathbf{p}_s)
\end{equation}
where we define the map $f_{\rm G}$ as the exact gate functional. Here we derive approximate gate functionals which evolve independently the SQPs of qubits acted on by single-qubit gates, and couple qubits acted on by two-qubit gates (here, CZ and CNOT). 
Analogous to DFT, where the electron interactions depend on the density, here we derive qubit-gate interactions that depend only on the SQPs, and use them to evolve the SQP vector. 
Recall that $p$ is the probability of measuring a single qubit in state $|1\rangle$. 
We define a single-qubit mean-field state consistent with this SQP: 
\begin{equation}
    |p\pm \rangle = \sqrt{1-p}\,\, |0 \rangle \pm \! \sqrt{p}\,\, |1 \rangle
\label{eq:mfstate}
\end{equation}
where we use $\pm$ to take into account two opposite phases between the $|0 \rangle$ and $|1 \rangle$ states.
\\
\indent 
For single-qubit gates, we apply the gate $U$ to this mean-field state, and then compute the probability of measuring $|1\rangle$ while taking the phase average over the $\pm$ states.
This approach provides explicit rules to update the SQPs at each step: 
\begin{equation} 
\label{eq:LPA}
  p_{s+1} = \frac{1}{2} \sum_{\pm}|\langle 1 | U |p_s\pm \rangle|^2 \,.
\end{equation}
This equation defines the local-probability approximation (LPA) gate functional. 
Using eq~\ref{eq:LPA}, we derive the following LPA update rules for common single-qubit gates:
\begin{align} 
&\text{Pauli X and Y:}  \hspace{18pt}p_{s+1} = 1 - p_s \nonumber\\ 
&\text{Pauli Z, S and T:}   \hspace{8pt} p_{s+1} = p_s \\
&\text{H, $\sqrt{X}$ and $\sqrt{Y}$:}  
\hspace{12pt} p_{s+1} = 0.5. \nonumber
\end{align} 
These results show that in our mean-field approach the Pauli X and Y gates flip the SQP, the Pauli Z, S and T gates leave the SQP unchanged as they act only on the phase, and the Hadamard, Pauli $\sqrt{X}$ and $\sqrt{Y}$ gates set the SQP to $1/2$.
\\
\indent
For the two-qubit gates considered here, CZ and CNOT, we use our intuition combined with the LPA rules to approximate the SQP evolution. In our probability-based formulation, the controlled unitary acts on the target qubit when $p^{(c)}>0.5$, namely when the control qubit is \lq\lq more one than zero\rq\rq. The probability $p^{(c)}$ of the control qubit is left unchanged and the probability $p^{(t)}$ of the target qubit is evolved according to the respective gate:
\begin{align}
\label{eq:LPA-last}
&\text{CZ: }\hspace{20pt} p^{(t)}_{s+1} = p^{(t)}_{s} \nonumber\\
&\text{CNOT:   if }\hspace{3pt} p^{(c)} < 0.5\text{\,,\,\,\,\,\,} 
p^{(t)}_{s+1} = p^{(t)}_{s}\\
&\hspace{44pt} \text{if  } p^{(c)} > 0.5\text{\,,\,\,\,\,\,} p^{(t)}_{s+1} = 1 - p^{(t)}_{s}\,\,. \nonumber
\end{align} 
The CZ result follows from the fact that our approximate Pauli Z gate leaves the SQP unchanged. 
The CNOT gate uses a $p^{(c)}=0.5$ threshold for controlling the target qubit, but the case $p^{(c)}=0.5$ is more subtle and needs a separate update rule: 
\begin{align}
\label{eq:lpacnot}
    &\text{CNOT:   if }\hspace{3pt} p^{(c)} = 0.5~\text{and}~p^{(t)} \!=\! 0~\text{or}~1\text{\,,\,\,\,\,\,} 
p^{(t)}_{s+1} = 0.5\\
    &\text{~~~~~~~~~~~if }\hspace{3pt} p^{(c)} = 0.5~\text{and}~p^{(t)} \!\neq\! 0~\text{or}~1\text{\,,\,\,\,\,\,} 
p^{(t)}_{s+1} =  1 - p^{(t)}_{s}\,\,. \nonumber
\end{align}
This choice allows us to address the important case of a Hadamard gate acting on the control qubit of a CNOT gate, as in the Bell-state preparation QC~\cite{Hidari}, a key building block in the random QCs discussed below. In particular, our CNOT and Hadamard update rules give the correct SQPs for all possible two-qubit initial basis states in the Bell-state preparation QC~\cite{Hidari} (see Table~\ref{tab:SI-tab1}). Additional discussion of gate functionals is provided below in Section 8.
\\
\indent
\begin{table}[h]
\begin{center}
\begin{minipage}{\textwidth}
\caption{\textbf{LPA functional applied to the Bell-state preparation QC}. Exact wave function $\Psi_s\,$, and the corresponding SQP vector $\mathbf{p}_s$, given as a function of step $s$ for the Bell-state preparation two-qubit QC~\cite{Hidari}. This QC consists of H applied to qubit 0 (step 1) followed by CNOT with control qubit 0 and target qubit 1 (step 2). As one can verify, the LPA rules in eqs~\ref{eq:LPA}$-$\ref{eq:LPA-last} give the same SQPs as the exact ones shown in the table, at all steps and for all initial states in the computational basis.}\label{tab:SI-tab1}
\begin{tabular*}{\textwidth}
{@{\extracolsep{\fill}}lccccc@{\extracolsep{\fill}}}
\hline
\vspace{1pt}
Initial state $\Psi_{s=0}$ & $\mathbf{p}_{s=0}$ & $\Psi_{s=1}$ & $\mathbf{p}_{s=1}$ & $\Psi_{s=2}$ & $\mathbf{p}_{s=2}$\\
\hline \vspace{-6pt}\\
\hspace{30pt}$\ket{00}$   & (0,0) & $\frac{1}{\sqrt{2}}\, (\ket{00} + \ket{10})$ & (0.5, 0) & $\frac{1}{\sqrt{2}}\, (\ket{00} + \ket{11})$  & (0.5, 0.5)\\
\hspace{30pt}$\ket{01}$   & (0,1)  & $\frac{1}{\sqrt{2}}\, (\ket{01} + \ket{11})$  & (0.5, 1) & $\frac{1}{\sqrt{2}}\, (\ket{01} + \ket{10})$  & (0.5, 0.5)\\
\hspace{30pt}$\ket{10}$   & (1,0)  & $\frac{1}{\sqrt{2}}\, (\ket{00} - \ket{10})$  & (0.5, 0) & $\frac{1}{\sqrt{2}}\, (\ket{00} - \ket{11})$ & (0.5, 0.5)\\
\hspace{30pt}$\ket{11}$   & (1,1)  & $\frac{1}{\sqrt{2}}\, (\ket{01} - \ket{11})$  & (0.5, 1) & $\frac{1}{\sqrt{2}}\, (\ket{01} - \ket{10})$ & (0.5, 0.5)\vspace{4pt}\\
\hline
\end{tabular*}
\end{minipage}
\end{center}
\end{table}

We implement these QC-DFT simulations using an in-house code (see Data Availability), and apply them to random and nonrandom QCs ranging from small to large, with up to a billion interacting qubits. 
For small QCs with less than $\sim$30 qubits, we compare the approximate SQPs with exact values obtained from wave-function (also known as state-vector) QC simulations carried out using the \textsc{QuEST} code~\cite{quest} (see Appendix). 
For this comparison, we define the 
\textit{SQP accuracy} $A_s$ as the fraction of SQPs predicted correctly by QC-DFT at \mbox{step $s$} (equivalently, the SQP error $1-A_s$ is the fraction of qubits with incorrect SQPs). 

\section{3. Random QC Simulations with the LPA Functional}
\begin{figure}[!ht]
\includegraphics[width=1.00\columnwidth]{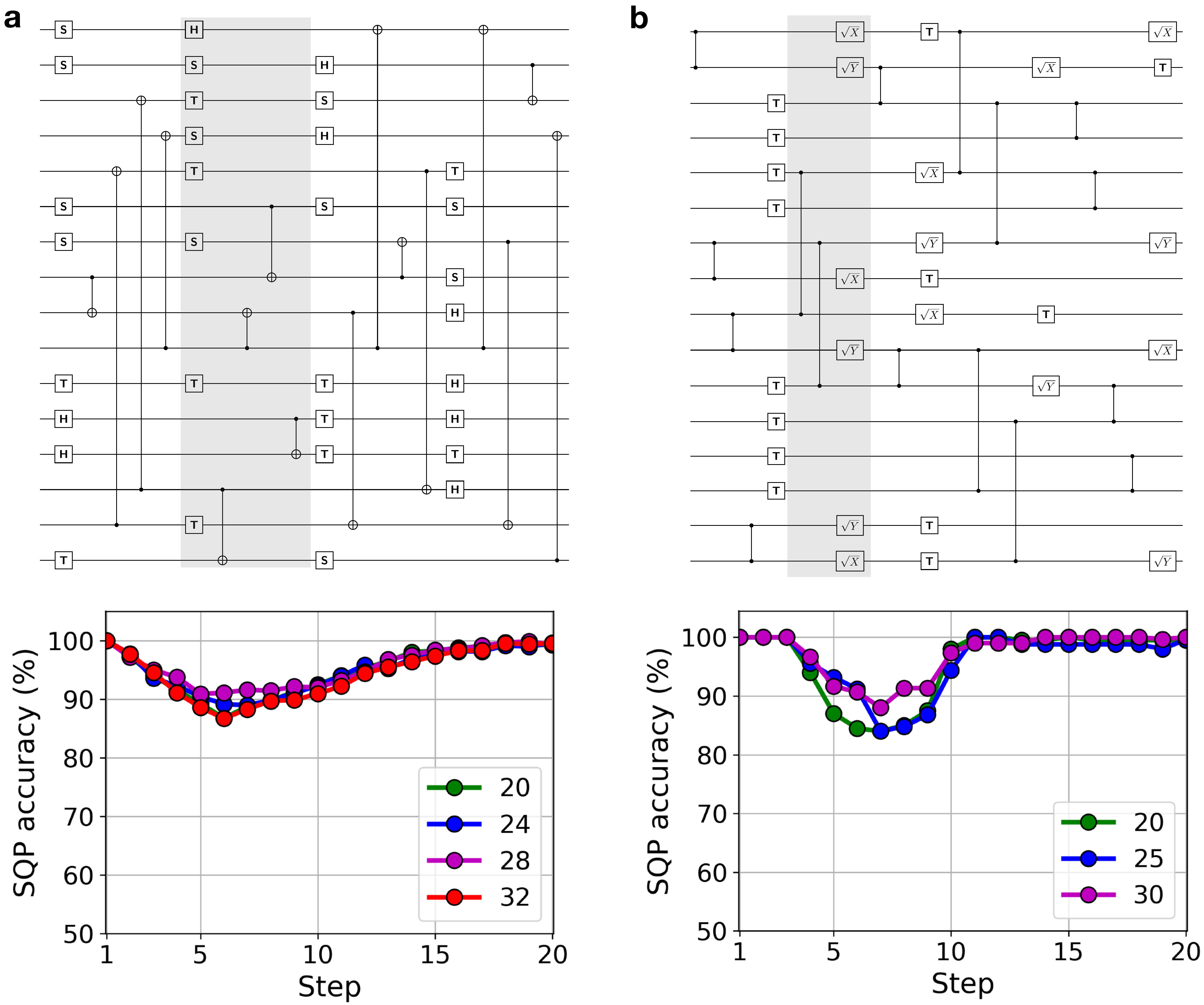}
\caption{\textbf{QC-DFT simulations of random quantum circuits using LPA gate functionals.} \textbf{a,} Random QC using a universal Clifford+T gate set. Each step consists of randomly chosen single-qubit gates applied to half of the qubits and CNOT gates applied to the other half (top). The accuracy of the simulated SQPs at each step is shown for this type of random QCs with different numbers of qubits (bottom). 
\textbf{b,} Random QC with T, $\sqrt{X}$, $\sqrt{Y}$ and CZ gates, taken from Ref.~\cite{Boixo} but with the H gates removed. Similar to (\textbf{a}), we plot the SQP accuracy for different QC sizes (bottom). Both types of QCs have a depth of 20 steps, with a single step shown in shaded gray. 
The SQPs are obtained by averaging results from 20 distinct random QC instances in (\textbf{a}) and 10 instances in (\textbf{b}). 
}\label{fig:QCs}
\end{figure}

First, we discuss results for random QCs with a universal Clifford+T gate set~\cite{clifft}, a moderate depth (20 steps), and QC sizes ranging from 20 to 32 qubits. In these QCs, at each step half of the qubits, chosen at random, are acted on by a randomly chosen single-qubit gate in the set, while the other half are acted on by CNOT gates that couple randomly-selected control and target qubits (Fig.~\ref{fig:QCs}a). 
In the exact simulations, the state-vector of the QC is initially set to $\ket{00\ldots0}_N$ and then evolved according to the gate sequence, while the exact SQPs are computed at each step via eq~\ref{eq:SQPdef}. The exact SQPs, with initial value of $\mathbf{p}=0$, evolve nontrivially for 10$-$15 steps, after which in our random QCs they reach a fully randomized value of $\mathbf{p}=0.5$. 
Our approximate QC-DFT simulations aim to capture the nontrivial SQP dynamics in the first 10$-$15 steps before randomization occurs.
\\
\indent
%
%
The results of our QC-DFT simulations for these Clifford+T QCs are shown in Fig.~\ref{fig:QCs}a. We find that our approach can predict the SQPs with an accuracy greater than $\sim$90\% at all steps. 
The highest error occurs near steps 5$-$7, where the qubits become nontrivially correlated, and then decreases to zero when the QC becomes fully randomized, with all SQPs trivially equal to $0.5$. 
Throughout the dynamics, the exact SQP values for most qubits are 0, 0.5, or 1 due to the combined action of the Hadamard and CNOT gates; therefore, the main challenge for the approximate simulations is to capture the transitions between these values, as further discussed below. 
We also simulate Clifford+T QCs with a different structure, which alternates one step where all qubits are acted on by randomly chosen single-qubit gates, and one step where all qubits are acted on by CNOT gates with randomly selected control and target qubits. The results, given in Fig.~S1 in the Supplementary Information, show a similar SQP accuracy of $90$\% or higher at all steps. 
\\
\indent
To demonstrate the versatility of our approach, we also simulate a different family of random QCs, introduced by Boixo et al.~\cite{Boixo}, which use a T, Pauli $\sqrt{X}$ and $\sqrt{Y}$, and CZ gate set (Fig.~\ref{fig:QCs}b). (We removed the Hadamard gate layer from the QCs in Ref.~\cite{Boixo} because it would make the SQPs trivially equal to 0.5 at all steps). 
The initial conditions are the same as in the Clifford+T circuits discussed above, but the quantum dynamics is richer, with more possible SQP values than in the Clifford+T case due to the combined effects of the T and square-root Pauli gates. Despite this greater complexity, our QC-DFT approach can simulate the dynamics of these QCs with an SQP accuracy greater than 85$-$90\% for sizes ranging from 20 to 30 qubits (Fig.~\ref{fig:QCs}b).
These results demonstrate that our QC-DFT approach, combined with the LPA rules, can accurately predict the SQPs for various random QCs without computing the exponentially complex QC wave function. 
\section{4. Improved Functionals: Multi-Gate Approximation}
We study whether the accuracy of QC-DFT can be improved by fine-tuning the gate functionals, in a spirit similar to improving electronic exchange-correlation functionals in DFT~\cite{Burke}. 
Many SQP errors in the LPA simulations derive from applying two consecutive times the same gate to a given qubit $-$ a situation analogous to a strong local interaction in the many-electron problem $-$ or from specific gate sequences acting on a qubit. 
Because our LPA focuses on local qubit-gate interactions at the current step, it lacks memory effects and cannot accurately describe such multi-gate correlations. 
To address this problem, we formulate multi-gate approximation (MGA) functionals encoding the effects of gate sequences, and apply them as a \textit{correction} to the LPA in the first $\sim$10 steps, where multi-gate correlations are important for predicting the SQP dynamics in our random QCs. 
\\
\indent
We define MGA-$n$ gate-functionals which treat explicitly single-qubit gate sequences with length $l\le n$. Their SQP update rules can be written as
\begin{equation}
\label{eq:MGA}
    p_{s+1} = |\langle 1 | \prod_{i=s-l}^{s}\!U_i\, |0 \rangle|^2,
\end{equation}
where $U_i$ is a single-qubit gate acting at step $i$.  
This approach captures the effect of sequences of $l$ gates, from step $s-l$ to the current step $s$, 
and focuses on early multi-gate corrections in the QC by assuming that the gate sequences act on the initial single-qubit state $\ket{0}$. When using these MGA-$n$ gate-functionals, the SQPs are evolved at each step using the LPA, but gate-sequences with length up to $n$ are searched at each step; when a sequence included in the functional is found, the SQP is updated according to eq~\ref{eq:MGA}. 
For example, for an MGA functional encoding a sequence of two Hadamard gates $U_H$, the first gate gives $p_s = 0.5$ due to the LPA rules, and the second gives $p_{s+1} = \langle 1 | U_H^2 |0\rangle = 0$, thus correcting the erroneous LPA value $p_{s+1} = 0.5$. Similarly, an MGA treating a sequence of two square-root of Pauli X gates gives $p_s = 0.5$ after the first and $p_{s+1} = \langle 1 | (\sqrt{\sigma_X})^2 |0\rangle = 1$ after the second $\sqrt{X}$ gate. 
\\
\indent 
We develop several MGA functionals (see Appendix) to improve the SQP accuracy by addressing the limitations of the LPA in our random QCs. 
For the Clifford+T QCs in Fig.~\ref{fig:QCs}a, our analysis of the LPA results reveals that two main gate sequences lead to SQP errors: the H$-$H sequence consisting of two consecutive Hadamard gates applied to the same qubit, which leads to $p_{s+1} = 0.5$ in the LPA instead of the exact $p_{s+1} = 0$, and the three-gate sequence H$-$T$-$H, which gives $p_{s+1} = 0.5$ in the LPA instead of the exact result $p_{s+1} = 0.146447$. 
Accordingly, we develop a simple MGA-3 functional addressing these two sequences, and apply it to our Clifford+T random QCs. 
Figure~\ref{fig:functionals}a compares the accuracy of this MGA-3 functional with the LPA for the random Clifford+T QCs in Fig.~\ref{fig:QCs}a. 
We apply the multi-gate corrections in the first 7 steps, and find a significant improvement of SQP accuracy, by roughly 5$-$8\%, during those steps. Beyond step $\sim$10, the QC state randomizes and the SQP accuracy becomes nearly identical for the two functionals. 
We find similar accuracy improvements for Clifford+T random QCs with a different structure, as shown in Fig.~S2 in Supplementary Information. 
\\
\indent
\begin{figure}[!t]
\includegraphics[width=1.0\columnwidth]{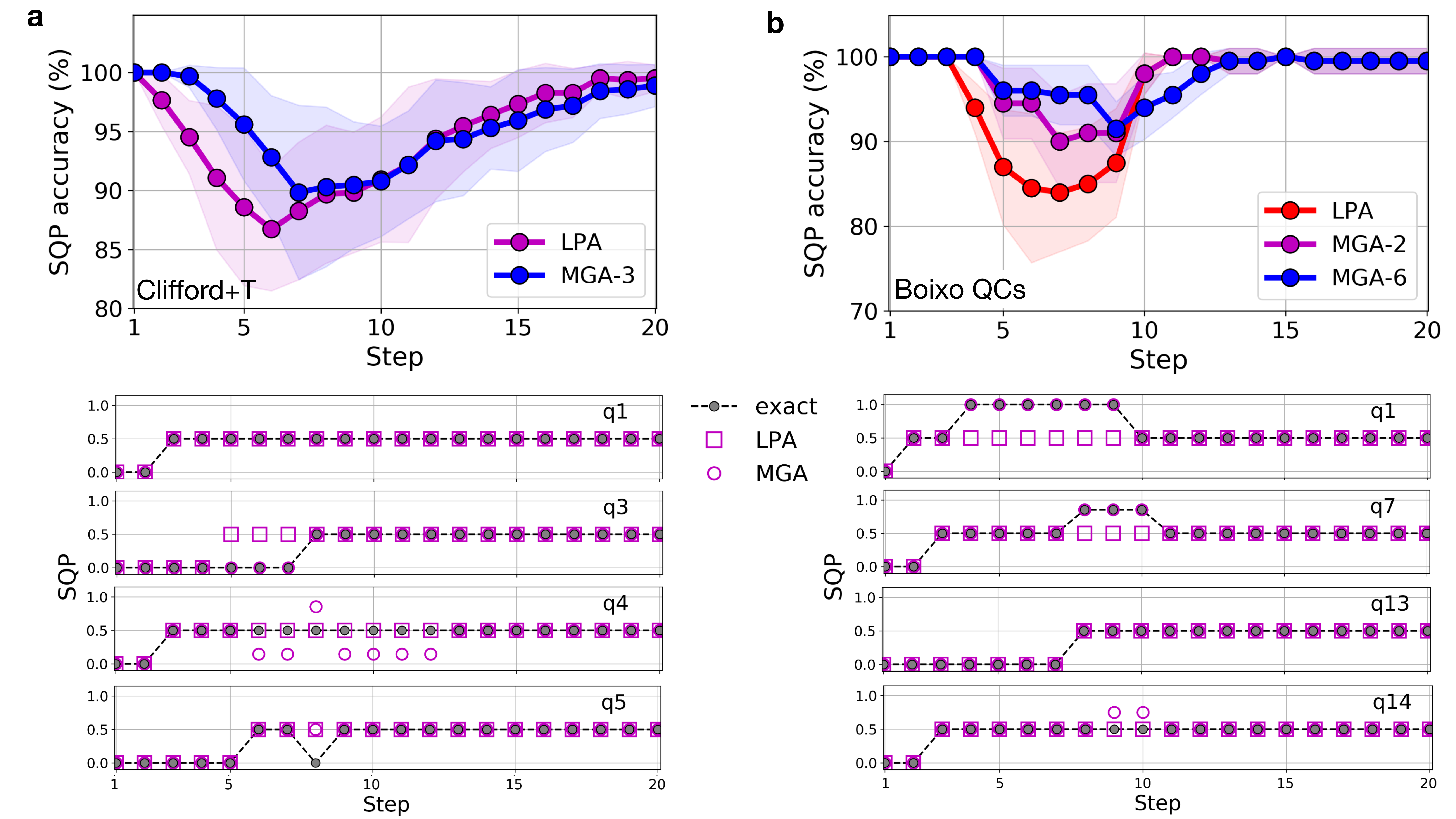}
\caption{\textbf{Optimized MGA gate functionals.} \textbf{a,} Accuracy comparison between the LPA and MGA-3 functionals applied to the Clifford+T circuits in Fig.~\ref{fig:QCs}a (top). The SQPs at each step are plotted for selected qubits for both functionals and compared with exact results (bottom). \textbf{b,} Accuracy comparison between the LPA and two different MGAs, MGA-2 and MGA-6, encoding respectively up to two- and six-gate sequences (top), shown together with the SQPs at each step for selected qubits (bottom). The results in (\textbf{a}) are for QCs with 32 qubits and in (\textbf{b}) for QCs with 20 qubits. The standard deviation of the SQP accuracy is shown for each curve using shaded colors. These results are obtained by averaging over the same number of QCs as in Fig.~\ref{fig:QCs}. 
}\label{fig:functionals}
\end{figure}
For the second family of random QCs discussed above, which employ T, Pauli $\sqrt{X}$ and $\sqrt{Y}$, and CZ gates, we develop two types of MGA functionals: MGA-2 addressing only sequences of two consecutive $\sqrt{X}$ or $\sqrt{Y}$ Pauli gates, and a systematically improved MGA-6 functional encoding sequences of up to six single- and two-qubit gates.
Both of these MGA functionals lead to accuracy improvements over the LPA, with the MGA-6 further improving over the simpler MGA-2 (Fig.~\ref{fig:functionals}b). 
For both types of random QCs studied here, we analyze the SQP dynamics for selected qubits (bottom panels in Fig.~\ref{fig:functionals}). We find that multi-gate corrections can have several different effects: the MGA can leave the LPA results unchanged, correct SQP errors in the LPA, fail to correct the LPA errors, or occasionally introduce errors not present in the LPA. 
When the MGA correction is successful, the SQP accuracy improvement typically lingers for several steps, leading to sizable accuracy improvements relative to the LPA. In addition, the multi-gate corrections allow us to capture SQP values different from 0, 0.5 and 1, the only possible values in the LPA.  
These results demonstrate a systematic approach for improving QC-DFT gate functionals by introducing memory effects and explicitly addressing multi-gate correlations.\\ 

\begin{figure}[!ht]
\includegraphics[width=0.9\columnwidth]{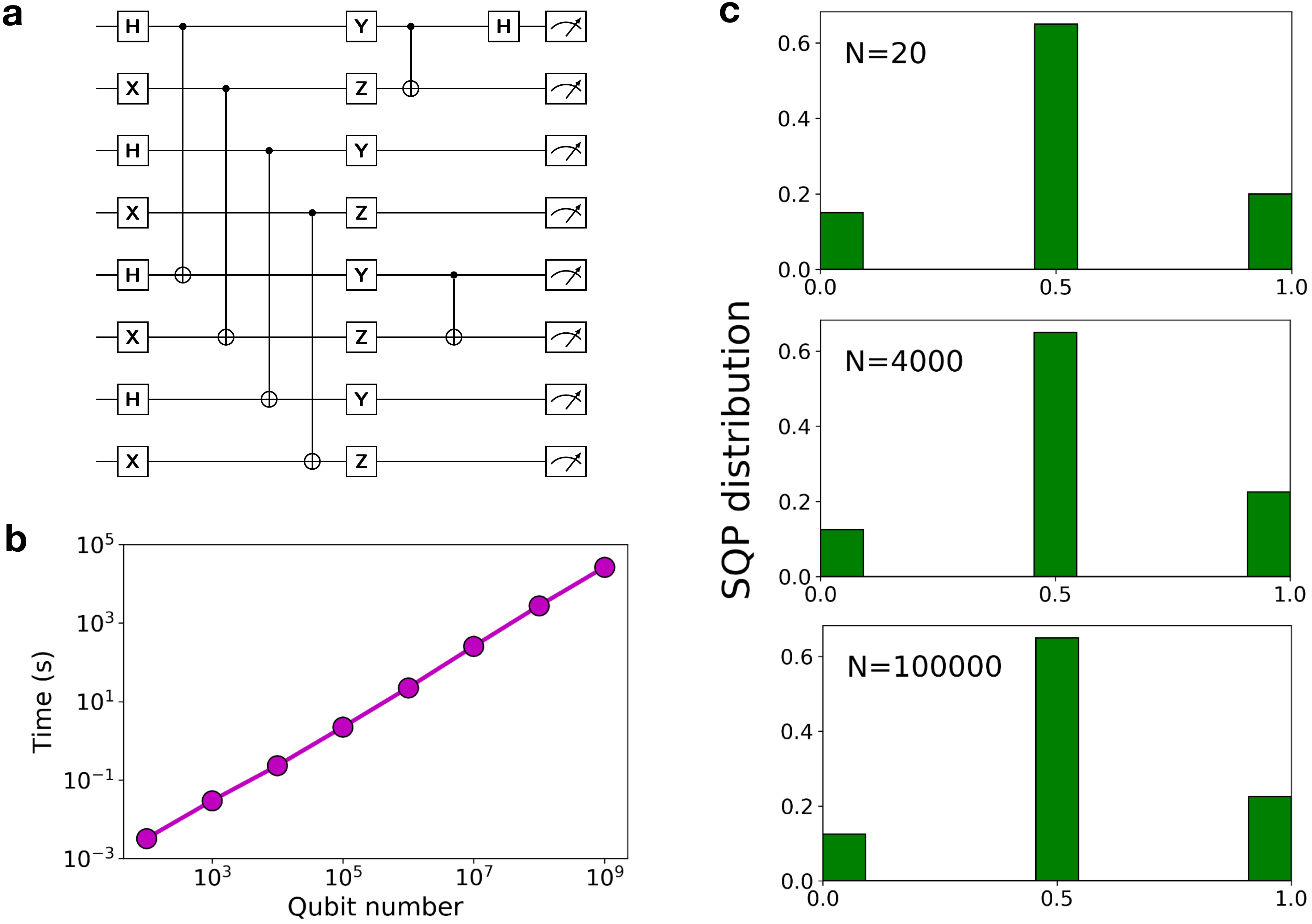}
\caption{\textbf{Computational cost and SQP distribution scaling with quantum \mbox{circuit} size.} \textbf{a,} Quantum circuit structure used to obtain the computational cost and SQP distribution as a function of QC size. The rules used to generate this type of QC are given in Appendix. \textbf{b,} Linear scaling of QC-DFT computation time with qubit number. \textbf{c,} Invariance of the SQP distribution with respect to QC size for QCs with the same structure, which is given in (\textbf{a}). Results  are shown for QCs with sizes of 20, 4000, and 10$^5$ qubits.
}\label{fig:scaling}
\end{figure}

\section{5. Large QCs and SQP Scaling}
The favorable scaling of our approach allows us to simulate very large QCs. We focus on a family of nonrandom QCs (with Hadamard, Pauli X, Y, Z, and CNOT gates), generated with a set of deterministic rules given in Appendix, whose circuit diagram for an example size of 8 qubits is shown in Fig.~\ref{fig:scaling}a. Using this class of QCs, with sizes ranging from 20 to 10$^9$ qubits, we demonstrate that the computational cost of QC-DFT has a linear scaling with number of qubits in the QC (Fig.~\ref{fig:scaling}b). We are able to complete the largest calculation, with size one billion qubits, using only a laptop computer for a few hours.
Note that both the memory and computational cost to obtain accurate SQPs scale linearly with QC size and depth in our QC-DFT approach,  
in clear contrast with the \textit{exact} SQPs from state-vector simulations~\cite{quest}, for which memory and computational cost scale exponentially with qubit number. 
\\
\indent 
Our analysis of the nonrandom QCs in Fig.~\ref{fig:scaling}a further reveals an intriguing physical result: for QCs with a given structure, the SQP distribution is independent of QC size, and thus is scale-invariant with respect to qubit number, as shown in Fig.~\ref{fig:scaling}c for three illustrative QC sizes. (Although the simulations for large $N$ values cannot be validated against exact results, we have verified that simulations for $N<30$ qubits achieve a 90\% SQP accuracy, similar to the other LPA results in this work.)
This finding shows that the SQP distribution is a fingerprint of the QC linked to its structure, a result reminiscent of the map between the electron density and the material structure in the Hohenberg-Kohn theorem of DFT~\cite{DFT}.   
Our analysis suggests that the SQPs are central quantities in mean-field simulations of QCs $-$ similar to the electron density in DFT, which is also a one-body marginal probability $-$ justifying the focus on SQPs in our approach.
\\

\section{6. Model Spin Hamiltonian}
We discuss an application of QC-DFT to obtain the ground-state energy of a \mbox{simple} spin Hamiltonian. 
Extending the variational quantum eigensolver (VQE)~\cite{VQE,VQE-2} discussion in Ref.~\cite{Hidari}, we model the $N$-qubit Hamiltonian $H = \bigotimes_{j=1}^N\,\sigma_z^{(j)}$, where $\sigma_z^{(j)}$ is the Pauli Z gate \mbox{acting} on qubit $j$, and search the energy minimum starting from the trial wave function $\ket{ \Psi(\boldsymbol{\theta}) } = \prod_{j=1}^N R_x^{(j)}(\theta_j) \ket{00\ldots0}_N$, where $R_x(\theta_j)$ is a rotation through angle $\theta_j$ around the $x$-axis applied to qubit $j$ (starting from an initial state $\ket{00\ldots0}_N$), and $\boldsymbol{\theta} = (\theta_1, \theta_2, \ldots, \theta_N)$ is the set of rotation angles parametrizing the wave function. 
For this example Hamiltonian, the energy $E(\boldsymbol{\theta})$ can be obtained analytically: the rotation $R_x(\theta)$ acting on each qubit gives a state $\ket{\varphi(\theta)}$ written as
\begin{equation}
R_x(\theta) \ket{0} \equiv \ket{\varphi(\theta)} = \cos (\theta/2) \ket{0} - i\, \sin (\theta/2) \ket{1},
\end{equation}
\vspace{-10pt}
and thus we obtain:\\
\begin{equation}
E(\boldsymbol{\theta}) = \braket{\Psi(\boldsymbol{\theta})|H |\Psi(\boldsymbol{\theta})} = \prod_{j=1}^N 
\braket{\varphi(\theta_j)| \sigma_z^{(j)} | \varphi(\theta_j)} = \prod_{j=1}^N [ \cos^2(\theta_j/2) - \sin^2(\theta_j/2)] = \prod_{j=1}^N \,\,\cos (\theta_j).
\label{eq:energy_VQE}
\end{equation}
\vspace{-10pt}
\\
\indent
In a state-vector simulation, preparing the trial wave function $\ket{ \Psi(\boldsymbol{\theta}) }$ and computing the associated energy $E(\boldsymbol{\theta})$ for any set of angles $\boldsymbol{\theta}$ requires the application of $N$ rotations about the $x$-axis, with a computational cost growing exponentially with qubit number $N$. 
Therefore, the search for the ground-state energy with state-vector simulations would require exponential resources. 
\\
\indent
Here we employ QC-DFT as an alternate route for efficient energy calculations. 
Using the LPA gate functional, the update rule for the $R_x(\theta)$ rotation is 
\begin{equation}
    p_{s+1} = \frac{1}{2}\sum_{\pm} \langle 1|R_x(\theta) | p_{s\,\pm} \rangle = p_s \cos^2(\theta/2) + (1-p_s) \sin^2(\theta/2)\,,
\end{equation}
which becomes $p_{s+1} = \sin^2(\theta/2)$ for our initial state with $p_s = 0$. 
Using as trial wave function the resulting mean-field state obtained with eq~\ref{eq:mfstate}, $\ket{p_{s+1}} = \cos (\theta/2) \ket{0} \pm \sin (\theta/2)\ket{1}$, we write the mean-field energy for a single-qubit Hamiltonian $\sigma_z$ as:
\begin{equation}
    \varepsilon_{\rm MF} (\theta,p_{s+1}) = \braket{p_{s+1} | \sigma_z | p_{s+1}} = \cos^2(\theta/2) - \sin^2(\theta/2) = \cos(\theta),
\end{equation}
where $\varepsilon_{\rm MF}$ depends explicitly on $p_{s+1}$, the SQP obtained after applying the $R_x(\theta)$ rotation in QC-DFT. 
This result can be extended to $N$ qubits, by applying rotations $R^{(j)}_x(\theta_j)$ to each qubit $j$ to obtain the $N$-qubit mean-field state $\ket{\mathbf{p}_{s+1}} = \ket{p^{(1)}_{s+1}, p^{(2)}_{s+1}, \ldots p^{(N)}_{s+1}}$, with $\ket{p^{(j)}_{s+1}} = \cos (\theta_j/2) \ket{0}_j \pm \sin (\theta_j/2)\ket{1}_j$ as above. 
For our Hamiltonian $H = \bigotimes_{j=1}^N\,\sigma_z^{(j)}$, the $N$-qubit mean-field energy $E_{\rm MF}$ factors into a product of single-qubit mean-field energies, and depends explicitly on the SQPs, $\mathbf{p}_{s+1}$, obtained after applying the $x$-axis rotation gates in QC-DFT: %
\begin{equation}
    E_{\rm MF} (\boldsymbol{\theta}, \mathbf{p}_{s+1}) = \braket{\mathbf{p}_{s+1} | H | \mathbf{p}_{s+1}} = \prod_{j=1}^N \varepsilon^{(j)}_{\rm MF} = \prod_{j=1}^N \cos(\theta_j).
\label{eq:mfenergy}
\end{equation}
This mean-field energy is identical to the exact analytic result in eq~\ref{eq:energy_VQE}, and it can be obtained in QC-DFT directly from the SQPs using approximate $R_x(\theta)$ rotations, without preparing the trial wave function. 
\begin{figure}[!t]
\includegraphics[width=0.7\columnwidth]{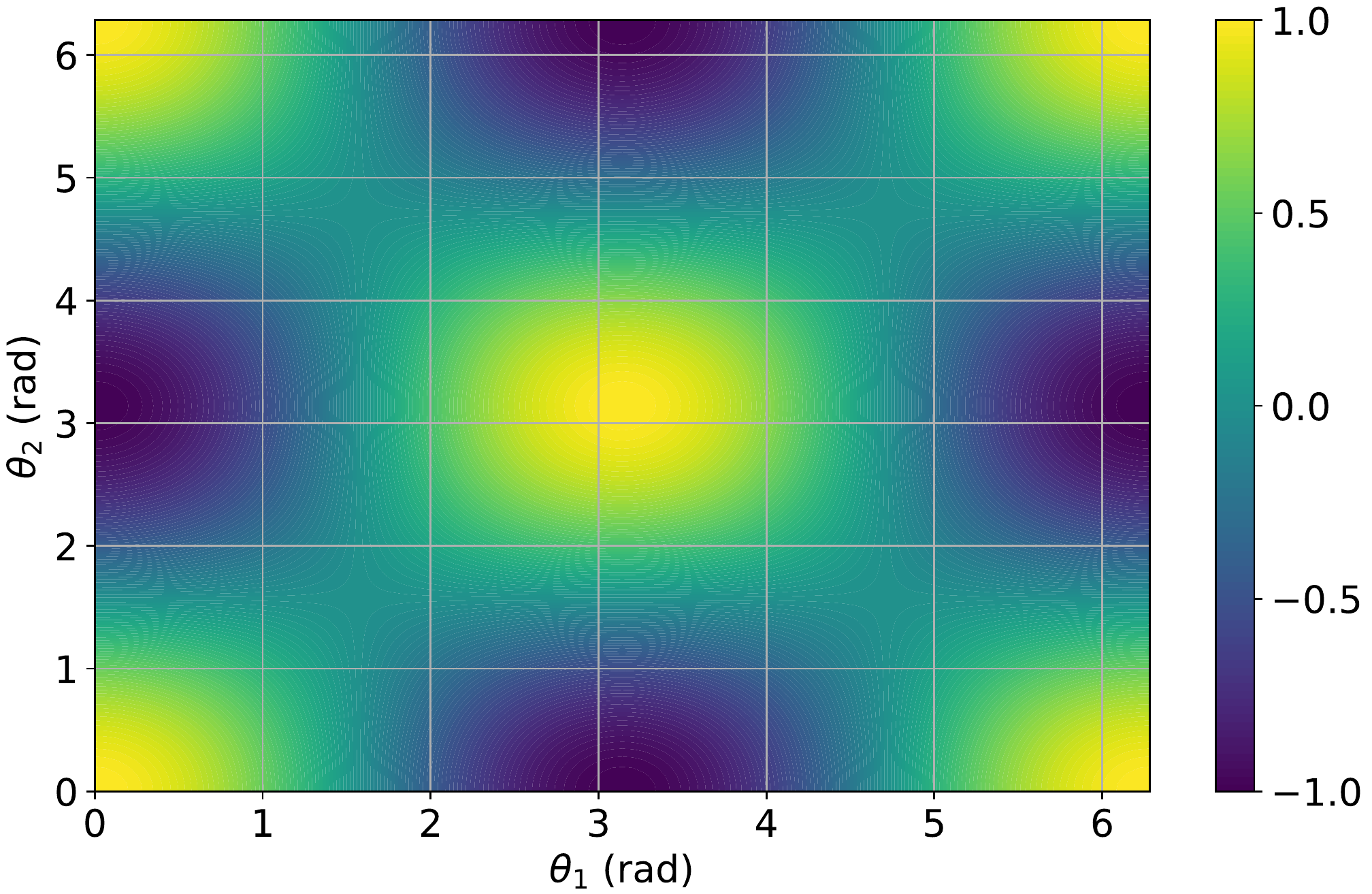}
\caption{\textbf{Energy calculation with QC-DFT.} Energy $E(\theta_1,\theta_2)$, color-coded in arbitrary units, for the two-qubit Hamiltonian $H=\sigma^{(1)}_z \sigma^{(2)}_z$ and the trial wave function $\Psi(\theta_1,\theta_2) = R^{(1)}_x(\theta_1)\, R^{(2)}_x(\theta_2) \ket{00}$. 
The energy is computed with QC-DFT on a fine grid of angles $(\theta_1,\theta_2)$ using eq~(\ref{eq:mfenergy}) together with the LPA gate functional for the rotations $R_x$.} 
\label{fig:vqe}
\end{figure}
Figure~\ref{fig:vqe} shows the energy $E(\theta_1,\theta_2)$ for the two-qubit case computed with QC-DFT on a fine grid of rotation angles $\theta_1$ and $\theta_2$. In this case, since $E(\theta_1,\theta_2) =  \cos(\theta_1)\cos(\theta_2)$, the minima are found for $(\theta_1,\theta_2) = (\pi,0)$ and $(0,\pi)$ (see Fig.~\ref{fig:vqe}). 
For the general $N$-qubit case, the energy $E(\boldsymbol{\theta})$ can be computed in QC-DFT using $\mathcal{O}(N)$ memory and computational resources by applying $x$-axis rotations in the LPA. 
Conversely, the same calculation has $\mathcal{O}(2^N)$ memory and computational cost in state-vector simulations. Although the example examined here has a simple analytic solution, it illustrates the point that QC-DFT may enable efficient mean-field calculations of ground-state energies. 
\section{7. Bernstein-Vazirani Quantum Algorithm and\\ Extension to Reduced Density Matrices}
\begin{figure}[!t]
\includegraphics[width=1.0\columnwidth]{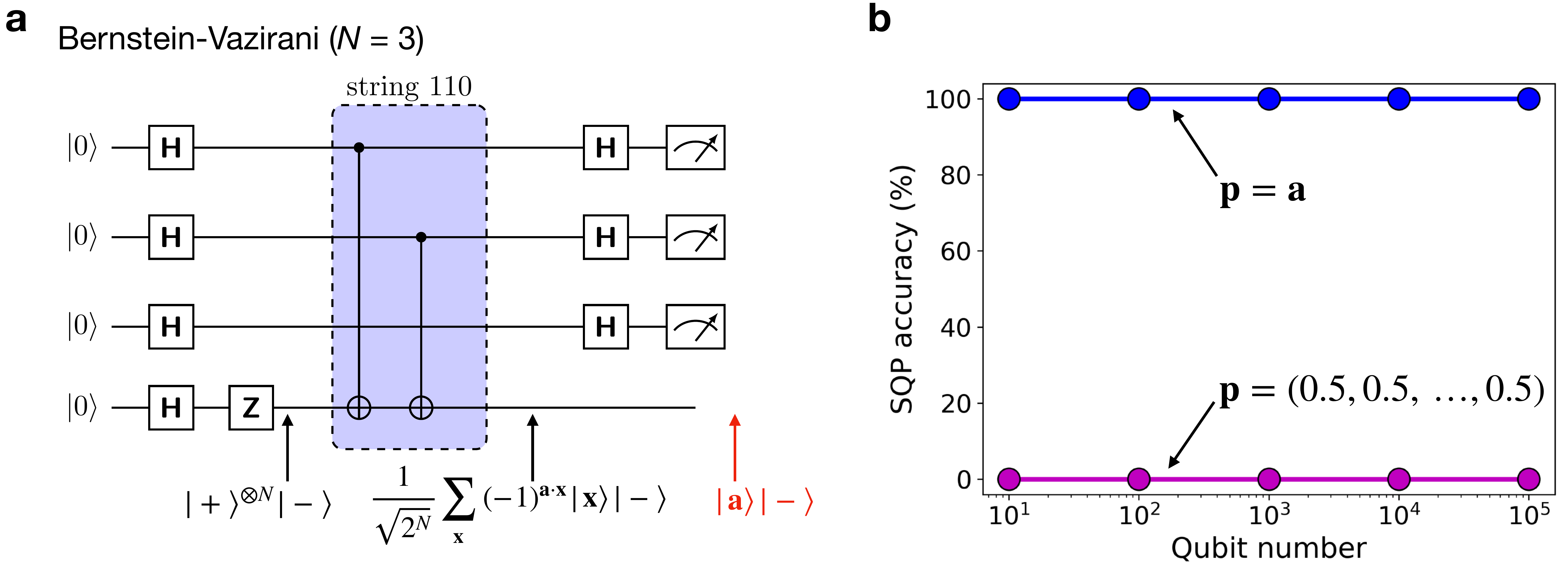}
\caption{\textbf{Simulations of the Bernstein-Vazirani quantum algorithm.} \textbf{a,} Schematic of the BV quantum circuit for $N\!=\!3$ qubits with secret string $\mathbf{a}=(110)$. The oracle, implemented as a set of CNOT gates, is shown as a dashed box with color shade, and the evolution of the exact state-vector is given at the bottom. \textbf{b,} SQP accuracy for QC-DFT simulations of the BV algorithm. The plot compares the accuracy of SQP-based simulations using the LPA rules (magenta) and simulations using 1-RDMs with the rules given in the text (blue). The calculations based on 1-RDMs achieve 100\% SQP accuracy and can correctly find the secret string $\mathbf{a}$, while LPA simulations incorrectly predict all SQPs as equal to 0.5. Results are shown for the QC in (\textbf{a}) with a number of qubits ranging from $N=10$ to $N=10^5$, in each case averaging over 100 randomly chosen secret strings.\vspace{15pt}\\}
\label{fig:BV}
\end{figure}

\vspace{-10pt}
Finally, we discuss the Bernstein-Vazirani (BV) quantum algorithm as a case study where SQP-based QC-DFT fails entirely. We also show that using the single-qubit reduced \mbox{density} matrix (1-RDM) as a different one-body quantity to formulate QC-DFT significantly improves the SQP accuracy, leading to results in perfect agreement with exact simulations. 
\\
\indent
We first show the failure of QC-DFT with LPA rules for the $N$-qubit BV quantum algorithm~\cite{Hidari}. In the BV quantum circuit, a secret string $\mathbf{a}$ is used in the oracle $f(\mathbf{x})=(\mathbf{a}\cdot \mathbf{x})~mod~2$ and encoded in the register using the phase kickback trick, which allows one to find the secret string in one query (as opposed to $N$ in the classical algorithm)~\cite{Hidari}. 
For this algorithm, QC-DFT predicts output SQPs all equal to 0.5, whereas the correct output SQPs equal (bitwise) the secret string $\mathbf{a}=(a_0\,  a_1 \ldots a_{N-1})$, where each $a_q$ is either 0 or 1. The reason for this failure can be understood by following the evolution of the SQPs in the BV circuit (see Fig.~\ref{fig:BV}a). The state used as input to the oracle is $\ket{\psi_0} = \ket{+}^{\otimes N} \ket{-}$, where  $\ket{+}^{\otimes N}$ is the state of the register and $\ket{-}$ the state of the ancilla qubit ($\ket{\pm}$ are Pauli X eigenstates). Therefore, all the SQPs are equal to 0.5 before the oracle is applied. 
The oracle is implemented as a set of CNOT gates applied between qubit $q$ and $N$ for qubits $q$ with $a_q=1$ in the string $\mathbf{a}=(a_0\,  a_1 \ldots a_{N-1})$. According to the LPA rules, if the SQPs of the control and target gates are both equal to 0.5, as is the case here, the SQPs will be left unchanged as they are both updated to 0.5 (see eq.~\ref{eq:lpacnot}). 
Following the application of the oracle, the state vector is $\ket{\psi_1} = \frac{1}{\sqrt{2^N}}\sum_{\mathbf{x}\in \{0,1\}^N} (-1)^{\mathbf{a}\cdot \mathbf{x}} \ket{\mathbf{x}} \ket{-}$. The final step applies a layer of Hadamard gates to the register, which transforms the register state so as to encode the secret string $\mathbf{a}$ bitwise. 
However, in QC-DFT the final layer of Hadamard gates sets the output SQPs to 0.5, in contrast with the correct values of 0 or 1 encoding the string $\mathbf{a}$. This leads to a zero SQP accuracy for QC-DFT with LPA rules in the case of the BV algorithm, as shown explicitly in Fig.~\ref{fig:BV}b. This result demonstrates that QC-DFT fails for the BV algorithm because using SQPs is insufficient to describe phase kickback. 
\\
\indent
We overcome these limitations by formulating a mean-field theory for a different one-body quantity, the single-qubit reduced density matrix (1-RDM). The 1-RDM for each qubit $n$ is defined as 
$\rho^{(n)} = \rm{Tr}_{\{j\}\ne n}(\ket{\Psi}\bra{\Psi})$, where $\ket{\Psi}$ is the QC state vector. 
The exact 1-RDMs can be computed by tracing out $N\!-\!1$ qubits from the full QC density matrix, $\ket{\Psi}\bra{\Psi}$, \mbox{obtained} from exact state-vector simulations. In contrast, in QC-DFT based on 1-RDMs the state-vector and full QC density matrix are never computed; rather, we update the 1-RDMs using mean-field update rules, similar to the SQP-based formulation discussed above. For 1-qubit unitary gates $U_{\rm G}$, the update rule $\rho_{s+1} = U_{\rm G}\, \rho_s\, U_{\rm G}^\dagger$ updates \textit{exactly} the 1-RDMs, as one can readily show. 
This removes any error in the simulation of 1-qubit gates, and addresses 1-qubit gates acting on the phase, including S, T, and Z which are neglected entirely in SQP-based QC-DFT. For the CNOT gate, we develop an approximate rule to update the 1-RDMs of the control and target qubits: 
\begin{align}
\label{eq:1RDM}
\rho^{(c)}_{s+1} &= \rm{Tr}_{t}[ U_{\rm CX} (\rho^{(c)}_{s} \otimes \rho^{(t)}_{s}) U_{\rm CX}^\dagger ]\\
\rho^{(t)}_{s+1} &= \rm{Tr}_{c}[ U_{\rm CX} (\rho^{(c)}_{s} \otimes \rho^{(t)}_{s}) U_{\rm CX}^\dagger ] \,\,, \nonumber
\vspace{-10pt}
\end{align}
where $U_{\rm CX}$ is the CNOT unitary matrix, while $\rm{Tr}_{c}$ and $\rm{Tr}_{t}$ indicate respectively tracing over the control and target qubits.  
These rules, which are motivated by our mean-field formulation, are exact if the control and target qubits are not entangled with each other.
\\
\indent
Figure~\ref{fig:BV}b shows that QC-DFT based on 1-RDMs, within the update rules given above, can correctly predict the SQPs for the BV algorithm, achieving a 100\% SQP accuracy. This result demonstrates that phase kickback, which is missed entirely in SQP-based QC-DFT, can be modeled correctly by extending QC-DFT to 1-RDMs. 
This result is promising because many quantum algorithms rely on phase kickback. Note also that the 1-RDM variant of QC-DFT has a computational cost that scales linearly with qubit number similar to the SQP-based variant. Such QC-DFT methods employing reduced density matrices will be explored more extensively in future work.

\section{8. Discussion}
\label{sec:discussion}
\vspace{-10pt}
\textbf{Gate functionals.}
We expand the discussion of gate functionals and  SQP update rules. Going from the state-vector ($2^N$ complex numbers) to the SQP (one real number) represents an enormous compression of information. Therefore, the map from state vectors to SQPs is clearly non-injective, and two or more different state vectors can share the same SQPs. For example, the two-qubit states $\ket{\psi_1} = \frac{1}{\sqrt{2}}\, (\ket{00}+\ket{11})$ and $\ket{\psi_2} = \frac{1}{\sqrt{2}}\, (\ket{01}+\ket{10})$ possess the same SQP vector, $\mathbf{p}_1=\mathbf{p}_2=(0.5,0.5)$. For these two states, applying a CNOT gate via the $\rm{CX}$ unitary gives $\ket{\psi'_1}=\rm{CX}(\ket{\psi_1}) =  \frac{1}{\sqrt{2}}\,  (\ket{00}+\ket{10})$ and $\ket{\psi'_2}= \rm{CX}(\ket{\psi_2}) = \frac{1}{\sqrt{2}}\,  (\ket{01}+\ket{11})$, with different SQP vectors of $\mathbf{p}'_1=(0.5,0)$ and $\mathbf{p}'_2=(0.5,1)$ respectively. 
Therefore, a single CNOT gate functional $\mathbf{p}'=f_G (\mathbf{p})$ that works exactly for both of these two-qubit states cannot exist, and clearly the situation is even more challenging for $N$-qubit states.
\\
\indent
More generally, the effect of 1- and 2-qubit gates on the SQPs is not universal and depends on the state vector the gate acts on. Therefore, an exact set of gate functionals that depends only on the SQPs cannot be derived because the map between SQPs and state vectors is not one-to-one. 
In contrast, QC-DFT employs \textit{approximate} gate functionals that depend only on the SQPs but not on the state vector (which is never computed), and applies them universally to all QCs. This approximation is consistent with the spirit of mean-field theories, where the interactions depend only on the one-body quantities (here, the SQPs) rather than on the unknown many-body wave function. 
The relation between SQPs, state vectors, unitary gates and approximate gate functionals is summarized in Fig.~\ref{fig:functionals}. 
\\
\indent
As an exact formulation of gate functionals is not possible, here we use physical intuition to derive approximate gate functionals and the corresponding update rules. For the 1-qubit gates, the LPA update rules are obtained by assuming that the qubits are independent (mean-field approximation) and that the only information available about the state of the QC is the SQP vector. 
For the CNOT gate, it is intuitive to control the target qubit only when the control-qubit SQP is greater than 0.5 (and thus it is \lq\lq more one than zero\rq\rq), but the case where the SQP of the control qubit is 0.5 is more subtle. For that case, we have adopted update rules that reproduce the SQPs for a specific QC that generates entanglement, the Bell-state preparation QC (see Table~\ref{tab:SI-tab1}). 
On this basis, it is quite remarkable that the LPA update rules, derived using simple intuition, work quite well, predicting SQPs with an accuracy higher than 85-90\% in a range of QCs as shown above. Note that if we chose an unphysical rule for the CNOT gate, or even a rule where the 0.5 control-qubit SQP case is not properly addressed, the SQP accuracy would drop significantly, resulting in highly inaccurate simulations.	More systematic ways of deriving accurate gate functionals would be desirable and will be explored in future work. 
\begin{figure}[!t]
\includegraphics[width=0.7\columnwidth]{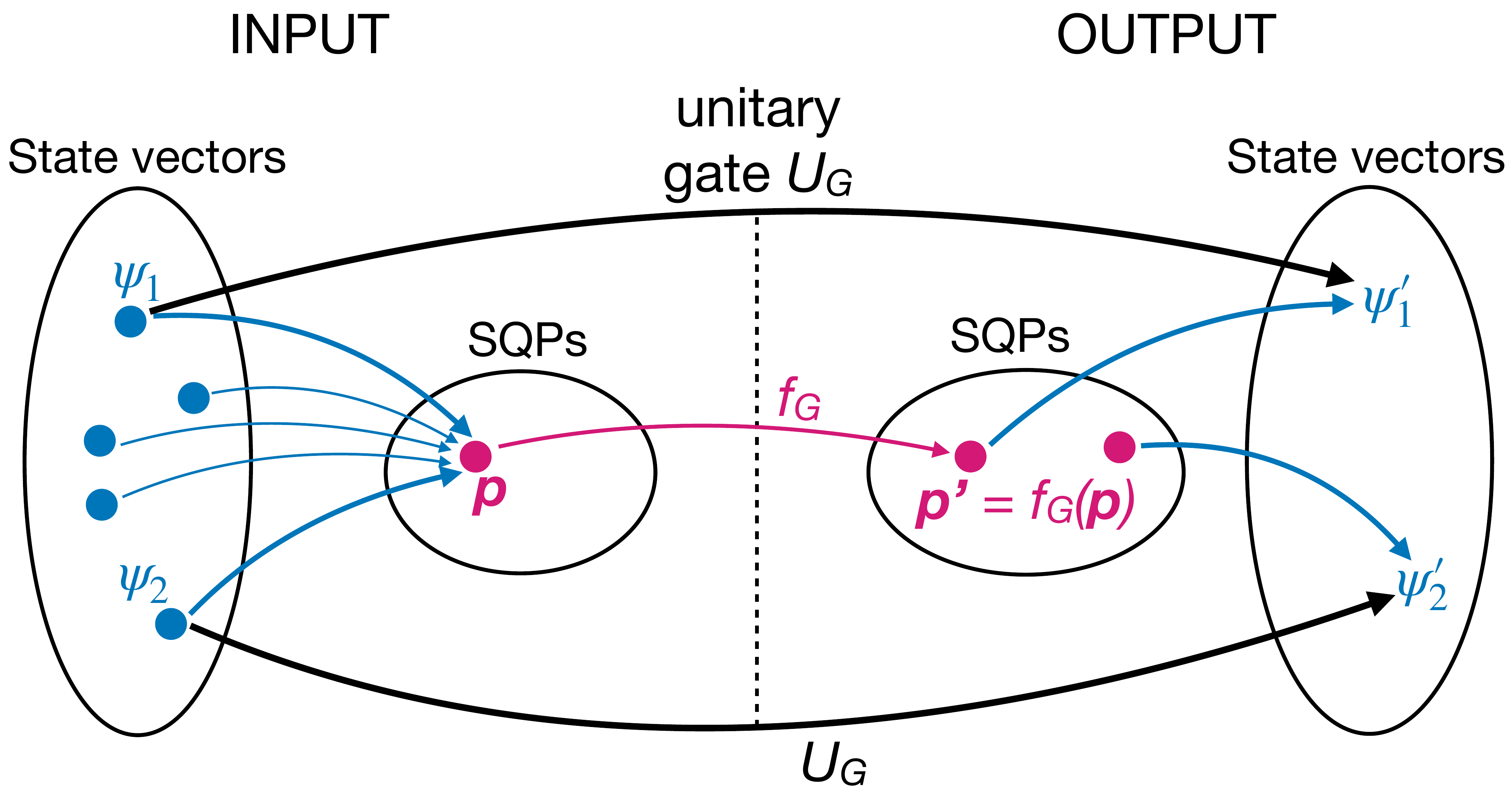}
\caption{\textbf{Schematic of QC-DFT and gate functionals.} The relations between state vectors, SQPs, gate functionals, and unitary gate transformations are illustrated. Multiple state-vectors $\psi_1$, $\psi_2$, etc. are all associated with the same SQP vector $\mathbf{p}$. The unitary gate maps these states respectively to $\psi'_1 = U_G (\psi_1$), $\psi'_2 = U_G (\psi_2$), etc., while the gate functional transforms the SQP vector to $\mathbf{p}'=f_G(\mathbf{p})$. The approximate gate functional works for $\psi_1$ because $\psi'_1$ corresponds to the SQP vector $\mathbf{p}'$. However, for the approximation fails for $\psi_2$ whose SQP vector is different from $\mathbf{p}'$. The information compression from state vectors to SQPs prevents the existence of universal gate functionals that work for arbitrary state vectors and QCs.}
\label{fig:functionals}
\end{figure}
\\
\textbf{Limitations and future directions.}
It is important to understand the limitations of the proposed QC-DFT approach. The SQP-based formulation of QC-DFT cannot capture qubit phase and interference effects, which are \mbox{essential} in quantum algorithms. Therefore, this method cannot compete with more established techniques such as tensor networks, and satisfactory results are expected mainly in QCs with low entanglement. For example, we have shown above that LPA simulations cannot describe the Bernstein-Vazirani quantum algorithm because the SQPs are insufficient to describe phase kickback. For the same reason, SQP-based QC-DFT also fails for the Deutsch-Jozsa and Grover algorithms~\cite{Hidari}. 
\\
\indent
Extensions of the QC-DFT formalism using reduced density matrices (RDMs) (instead of the SQPs), presented briefly above and inspired by recent advances in electronic structure methods using RDMs~\cite{mazziotti-1,Mazziotti-2}, enable an improved description of gates such as S, T, Pauli Z and controlled-Z, which act on the qubit phase and are ignored in our current SQP-based approach. For example, we have shown above that QC-DFT using 1-RDMs can accurately predict the SQPs in the BV quantum algorithm; such extensions of the QC-DFT method based on RDMs will be discussed more extensively elsewhere.
Our formulation of a DFT analog for QC simulations motivates several future research directions, including using machine learning to improve the QC-DFT gate functionals, as shown recently for exchange-correlation functionals in DFT~\cite{ML-DFT,Review-ML-DFT}, and applying QC-DFT and its RMD-based extensions to spin Hamiltonians and quantum algorithms.
\section{9. Conclusion}
In summary, we demonstrated mean-field simulations of QCs inspired by DFT. The approach shown in this work, called QC-DFT, can accurately predict the SQPs $-$ marginals of the full QC probability distribution $-$ with low computational cost (despite their formal exponential scaling) for various random and nonrandom QCs. 
Although the current approach is not generic and is limited to QCs with low entanglement, improvements to this formalism based on one- and two-qubit RDMs may enable simulations of more general classes of QCs.

\section{\hspace{-20pt} Appendix A. Numerical Methods}
\label{append:comp}
\vspace{-6pt}
\noindent 
\textbf{Exact QC simulations.} 
The exact QC simulations are carried out using the \textsc{QuEST} code~\cite{quest}. All single- and two-qubit gates are used as provided in the code. We use appropriate rotation operations to implement the square root Pauli gates, and compute the exact SQPs from the state vector. Example input files for \textsc{QuEST} are available in the data sets accompanying this manuscript.\vspace{6pt}
\\
\noindent
%
\textbf{Multi-gate functionals.}
The MGA-$n$ functionals are implemented in our QC-DFT code by looking for specific gate sequences in the QC. If a gate sequence encoded in the functional is found within cycle $s_{\rm max}$, the SQPs are updated using $p_{s+1}$ from eq~\ref{eq:MGA}. These multi-gate corrections are applied only up to once for each qubit. For the Clifford+T QCs, the MGA-3 functional used in Fig.~\ref{fig:functionals}a corrects for the gate sequences H-H (using $p_{s+1}=0$) and H-T-H ($p_{s+1}=0.146447$) up to cycle $s_{\rm max}=7$, including cases where CNOT gates act on the qubit within these sequences. 
This means that CNOT control and target operations are ignored when looking for these gate sequences $-$ for example, the gate sequence H$-$CNOT$-$H acting on a qubit is treated as H$-$H and corrected.  For the QCs in Fig.~\ref{fig:functionals}b, the MGA-2 functional corrects for the $\sqrt{X}$$-$$\sqrt{X}$ and $\sqrt{Y}$$-$$\sqrt{Y}$ sequences up to cycle 
$s_{\rm max}=7$. In this case, the CZ control operations are ignored when looking for gate sequences, while CZ target operations are taken into account. For example, if the gate sequence $\sqrt{X}$$-$CZ$-$$\sqrt{X}$ involves the CZ control qubit, it is treated as $\sqrt{X}$$-$$\sqrt{X}$ and the multi-gate correction is applied. If the same sequence is found for the CZ target qubit, the multi-gate correction is not applied. 
\\
\indent
The MGA-6 functional includes several multi-gate corrections with up to 6-gate sequences, applied up to cycle $s_{\rm max}$ between 6 and 10 depending on the sequence. Some sequences take into account CZ gates, while others ignore them. Next we provide the full list of gate-sequences for our MGA-6 functional, using a naming convention for gate sequences where, for a given qubit, the rightmost gate acts at the current step, and the leftmost gate acts at the earliest step in the sequence; steps where no gates act on the qubit are ignored. This means that sequences are given in the same order as when reading the QC from left to right, ignoring steps with no gates. The CZ gates are explicitly taken into account, in the same way for control and target qubits, unless otherwise stated. 
With these conventions, the gate sequences treated in our MGA-6 functional are as follows: 2-gate sequences
$\sqrt{X}$$-$$\sqrt{X}$ and $\sqrt{Y}$$-$$\sqrt{Y}$  (both with $p_{s+1} \!=\! 1$ and $s_{\rm max} \!=\! 6$); 3-gate sequences  $\sqrt{X}$$-$CZ$-$$\sqrt{X}$ and $\sqrt{Y}$$-$CZ$-$$\sqrt{Y}$ ($p_{s+1} \!=\! 1$, $s_{\rm max} \!=\! 6$), $\sqrt{Y}$$-$T$-$$\sqrt{X}$ ($p_{s+1} \!=\! 0.14645$, $s_{\rm max} \!=\! 8$, CZ gates ignored) and $\sqrt{Y}$$-$T$-$$\sqrt{Y}$, $\sqrt{X}$$-$T$-$$\sqrt{X}$, and $\sqrt{X}$$-$T$-$$\sqrt{Y}$ ($p_{s+1} \!=\! 0.85355$, $s_{\rm max} \!=\! 8$, CZ gates ignored); 4-gate sequences T$-$$\sqrt{Y}$$-$T$-$$\sqrt{X}$ and T$-$$\sqrt{X}$$-$T$-$$\sqrt{Y}$ ($p_{s+1} \!=\! 0.75$, $s_{\rm max} \!=\! 10$, CZ gates ignored); 5-gate
sequences CZ$-$$\sqrt{X}$$-$T$-$CZ$-$$\sqrt{X}$, CZ$-$$\sqrt{X}$$-$T$-$CZ$-$$\sqrt{Y}$, CZ$-$$\sqrt{Y}$$-$T$-$CZ$-$$\sqrt{X}$, and CZ$-$$\sqrt{Y}$$-$T$-$CZ$-$$\sqrt{Y}$ ($p_{s+1} \!=\! 0.5$, $s_{\rm max} \!=\! 10$); 6-gate sequences $\sqrt{Y}$$-$CZ$-$$\sqrt{X}$$-$T$-$CZ$-$$\sqrt{X}$ ($p_{s+1} \!=\! 0.14645$, $s_{\rm max} \!=\! 10$),   $\sqrt{X}$$-$CZ$-$$\sqrt{Y}$$-$T$-$CZ$-$$\sqrt{X}$ and $\sqrt{X}$$-$CZ$-$$\sqrt{Y}$$-$T$-$CZ$-$$\sqrt{Y}$ ($p_{s+1} \!=\! 0.85355$, $s_{\rm max} \!=\! 10$). 
These sequences can be found in the MGA-$n$ QC-DFT codes provided in the data sets accompanying this manuscript.\vspace{6pt}
\\
\noindent
\textbf{Nonrandom QCs.} The nonrandom QCs used for the scaling calculations in Fig.~\ref{fig:scaling} are generated using deterministic rules. For a QC with size $N$ qubits, step 1 consists of alternating H and Pauli X gates; in step 2, a CNOT gate connects each qubit $i < N/2$ (control) to qubit $i+N/2$ (target); step 3 consists of alternating Pauli Y and Z gates; step 4 has CNOT gates every 4 qubits, each with neighboring control and target qubits; step 5 applies H gates every 10 qubits. Only QCs with size $N$ multiple of 4 and 10 have the same structure, and thus give the same SQP distribution as shown above. Codes for generating these QCs and reproducing the calculations in Fig.~\ref{fig:scaling} are provided in the data sets accompanying this manuscript.
\newpage
\section*{\hspace{-16pt}Data Availability}
\vspace{-15pt}
The data sets generated and analyzed in this study, as well as the QC-DFT codes, will be made available in the CaltechDATA repository. Additional data and information are available upon reasonable request. 
The \textsc{QuEST} code~\cite{quest} used for the exact QC simulations is an open source software, which can be downloaded at \url{https://quest.qtechtheory.org}. The QC drawings were prepared using the Quantikz LaTeX package~\cite{tikz}, which can be downloaded at \url{https://ctan.org/pkg/quantikz}. The QC-DFT {\sc Python} code will be made available in the CaltechDATA repository. 
\vspace{-15pt}
\section*{\hspace{-16pt}Supporting Information}
\vspace{-15pt}
\noindent
The Supporting Information is available free of charge at [link]. Figure S1, additional LPA simulations of Clifford+T random QCs; Figure S2, additional results for optimized MGA functionals.

\vspace{-15pt}
\section*{\hspace{-16pt}Acknowledgements}
\vspace{-15pt}
The author acknowledges fruitful discussions with Sijing Du, Sandeep Sharma and \mbox{Garnet} Chan. This work was supported by the U.S. Department of Energy, Office of Science, Office of Advanced Scientific Computing Research and Office of Basic Energy Sciences, Scientific Discovery through Advanced Computing (SciDAC) program under Award Number DE-SC0022088. \vspace{-30pt}\\%
\newpage

\begin{mcitethebibliography}{37}
\providecommand*\natexlab[1]{#1}
\providecommand*\mciteSetBstSublistMode[1]{}
\providecommand*\mciteSetBstMaxWidthForm[2]{}
\providecommand*\mciteBstWouldAddEndPuncttrue
  {\def\EndOfBibitem{\unskip.}}
\providecommand*\mciteBstWouldAddEndPunctfalse
  {\let\EndOfBibitem\relax}
\providecommand*\mciteSetBstMidEndSepPunct[3]{}
\providecommand*\mciteSetBstSublistLabelBeginEnd[3]{}
\providecommand*\EndOfBibitem{}
\mciteSetBstSublistMode{f}
\mciteSetBstMaxWidthForm{subitem}{(\alph{mcitesubitemcount})}
\mciteSetBstSublistLabelBeginEnd
  {\mcitemaxwidthsubitemform\space}
  {\relax}
  {\relax}

\bibitem[Preskill(2018)]{NISQ}
Preskill,~J. Quantum Computing in the {NISQ} era and beyond. \emph{Quantum}
  \textbf{2018}, \emph{2}, 79\relax
\mciteBstWouldAddEndPuncttrue
\mciteSetBstMidEndSepPunct{\mcitedefaultmidpunct}
{\mcitedefaultendpunct}{\mcitedefaultseppunct}\relax
\EndOfBibitem
\bibitem[Nielsen and Chuang(2010)Nielsen, and Chuang]{Chuang}
Nielsen,~M.~A.; Chuang,~I.~L. \emph{Quantum Computation and Quantum
  Information: 10th Anniversary Edition}; Cambridge University Press,
  2010\relax
\mciteBstWouldAddEndPuncttrue
\mciteSetBstMidEndSepPunct{\mcitedefaultmidpunct}
{\mcitedefaultendpunct}{\mcitedefaultseppunct}\relax
\EndOfBibitem
\bibitem[Arute \latin{et~al.}(2019)Arute, Arya, Babbush, Bacon, Bardin,
  Barends, Biswas, Boixo, Brandao, Buell, Authors, Authors, Authors, Authors,
  Authors, and Authors]{Google}
Arute,~F. \latin{et~al.}  Quantum supremacy using a programmable
  superconducting processor. \emph{Nature} \textbf{2019}, \emph{574},
  505--510\relax
\mciteBstWouldAddEndPuncttrue
\mciteSetBstMidEndSepPunct{\mcitedefaultmidpunct}
{\mcitedefaultendpunct}{\mcitedefaultseppunct}\relax
\EndOfBibitem
\bibitem[De~Raedt \latin{et~al.}(2019)De~Raedt, Jin, Willsch, Willsch,
  Yoshioka, Ito, Yuan, and Michielsen]{juqs}
De~Raedt,~H.; Jin,~F.; Willsch,~D.; Willsch,~M.; Yoshioka,~N.; Ito,~N.;
  Yuan,~S.; Michielsen,~K. Massively parallel quantum computer simulator,
  eleven years later. \emph{Comput. Phys. Commun.} \textbf{2019}, \emph{237},
  47--61\relax
\mciteBstWouldAddEndPuncttrue
\mciteSetBstMidEndSepPunct{\mcitedefaultmidpunct}
{\mcitedefaultendpunct}{\mcitedefaultseppunct}\relax
\EndOfBibitem
\bibitem[Smelyanskiy \latin{et~al.}(2016)Smelyanskiy, Sawaya, and
  Aspuru-Guzik]{qhipster}
Smelyanskiy,~M.; Sawaya,~N. P.~D.; Aspuru-Guzik,~A. {qHiPSTER: The Quantum High
  Performance Software Testing Environment}. \emph{arXiv 1601.07195}
  \textbf{2016}, \relax
\mciteBstWouldAddEndPunctfalse
\mciteSetBstMidEndSepPunct{\mcitedefaultmidpunct}
{}{\mcitedefaultseppunct}\relax
\EndOfBibitem
\bibitem[Jones \latin{et~al.}(2019)Jones, Brown, Bush, and Benjamin]{quest}
Jones,~T.; Brown,~A.; Bush,~I.; Benjamin,~S.~C. {QuEST} and high performance
  simulation of quantum computers. \emph{Sci. Rep.} \textbf{2019}, \emph{9},
  1--11\relax
\mciteBstWouldAddEndPuncttrue
\mciteSetBstMidEndSepPunct{\mcitedefaultmidpunct}
{\mcitedefaultendpunct}{\mcitedefaultseppunct}\relax
\EndOfBibitem
\bibitem[Chen \latin{et~al.}(2018)Chen, Zhou, Xue, Yang, Guo, and Guo]{64qubit}
Chen,~Z.-Y.; Zhou,~Q.; Xue,~C.; Yang,~X.; Guo,~G.-C.; Guo,~G.-P. 64-qubit
  quantum circuit simulation. \emph{Sci. Bull.} \textbf{2018}, \emph{63},
  964--971\relax
\mciteBstWouldAddEndPuncttrue
\mciteSetBstMidEndSepPunct{\mcitedefaultmidpunct}
{\mcitedefaultendpunct}{\mcitedefaultseppunct}\relax
\EndOfBibitem
\bibitem[Jozsa(2006)]{Jozsa}
Jozsa,~R. On the simulation of quantum circuits. \emph{arXiv quant-ph/0603163}
  \textbf{2006}, \relax
\mciteBstWouldAddEndPunctfalse
\mciteSetBstMidEndSepPunct{\mcitedefaultmidpunct}
{}{\mcitedefaultseppunct}\relax
\EndOfBibitem
\bibitem[Markov and Shi(2008)Markov, and Shi]{Markov2008}
Markov,~I.~L.; Shi,~Y. Simulating Quantum Computation by Contracting Tensor
  Networks. \emph{{SIAM} J. Comput.} \textbf{2008}, \emph{38}, 963--981\relax
\mciteBstWouldAddEndPuncttrue
\mciteSetBstMidEndSepPunct{\mcitedefaultmidpunct}
{\mcitedefaultendpunct}{\mcitedefaultseppunct}\relax
\EndOfBibitem
\bibitem[Vidal(2003)]{Vidal}
Vidal,~G. Efficient Classical Simulation of Slightly Entangled Quantum
  Computations. \emph{Phys. Rev. Lett.} \textbf{2003}, \emph{91}, 147902\relax
\mciteBstWouldAddEndPuncttrue
\mciteSetBstMidEndSepPunct{\mcitedefaultmidpunct}
{\mcitedefaultendpunct}{\mcitedefaultseppunct}\relax
\EndOfBibitem
\bibitem[Yoran and Short(2006)Yoran, and Short]{Yoran}
Yoran,~N.; Short,~A.~J. Classical Simulation of Limited-Width Cluster-State
  Quantum Computation. \emph{Phys. Rev. Lett.} \textbf{2006}, \emph{96},
  170503\relax
\mciteBstWouldAddEndPuncttrue
\mciteSetBstMidEndSepPunct{\mcitedefaultmidpunct}
{\mcitedefaultendpunct}{\mcitedefaultseppunct}\relax
\EndOfBibitem
\bibitem[Jozsa and Miyake(2008)Jozsa, and Miyake]{Jozsa-2}
Jozsa,~R.; Miyake,~A. Matchgates and classical simulation of quantum circuits.
  \emph{Proceedings of the Royal Society A: Mathematical, Physical and
  Engineering Sciences} \textbf{2008}, \emph{464}, 3089--3106\relax
\mciteBstWouldAddEndPuncttrue
\mciteSetBstMidEndSepPunct{\mcitedefaultmidpunct}
{\mcitedefaultendpunct}{\mcitedefaultseppunct}\relax
\EndOfBibitem
\bibitem[Zhou \latin{et~al.}(2020)Zhou, Stoudenmire, and Waintal]{PRX}
Zhou,~Y.; Stoudenmire,~E.~M.; Waintal,~X. What Limits the Simulation of Quantum
  Computers? \emph{Phys. Rev. X} \textbf{2020}, \emph{10}, 041038\relax
\mciteBstWouldAddEndPuncttrue
\mciteSetBstMidEndSepPunct{\mcitedefaultmidpunct}
{\mcitedefaultendpunct}{\mcitedefaultseppunct}\relax
\EndOfBibitem
\bibitem[Gray(2018)]{quimb}
Gray,~J. quimb: A python package for quantum information and many-body
  calculations. \emph{J. Open Source Softw.} \textbf{2018}, \emph{3}, 819\relax
\mciteBstWouldAddEndPuncttrue
\mciteSetBstMidEndSepPunct{\mcitedefaultmidpunct}
{\mcitedefaultendpunct}{\mcitedefaultseppunct}\relax
\EndOfBibitem
\bibitem[Jónsson \latin{et~al.}(2018)Jónsson, Bauer, and Carleo]{Carleo}
Jónsson,~B.; Bauer,~B.; Carleo,~G. Neural-network states for the classical
  simulation of quantum computing. \emph{arXiv 1808.05232} \textbf{2018},
  \relax
\mciteBstWouldAddEndPunctfalse
\mciteSetBstMidEndSepPunct{\mcitedefaultmidpunct}
{}{\mcitedefaultseppunct}\relax
\EndOfBibitem
\bibitem[Martin \latin{et~al.}(2016)Martin, Reining, and Ceperley]{Reining}
Martin,~R.~M.; Reining,~L.; Ceperley,~D.~M. \emph{Interacting Electrons: Theory
  and Computational Approaches}; Cambridge University Press, 2016\relax
\mciteBstWouldAddEndPuncttrue
\mciteSetBstMidEndSepPunct{\mcitedefaultmidpunct}
{\mcitedefaultendpunct}{\mcitedefaultseppunct}\relax
\EndOfBibitem
\bibitem[Martin(2020)]{DFT}
Martin,~R.~M. \emph{Electronic Structure: Basic Theory and Practical Methods};
  Cambridge University Press, 2020\relax
\mciteBstWouldAddEndPuncttrue
\mciteSetBstMidEndSepPunct{\mcitedefaultmidpunct}
{\mcitedefaultendpunct}{\mcitedefaultseppunct}\relax
\EndOfBibitem
\bibitem[Car and Parrinello(1985)Car, and Parrinello]{CP}
Car,~R.; Parrinello,~M. Unified Approach for Molecular Dynamics and
  Density-Functional Theory. \emph{Phys. Rev. Lett.} \textbf{1985}, \emph{55},
  2471--2474\relax
\mciteBstWouldAddEndPuncttrue
\mciteSetBstMidEndSepPunct{\mcitedefaultmidpunct}
{\mcitedefaultendpunct}{\mcitedefaultseppunct}\relax
\EndOfBibitem
\bibitem[Ceperley and Alder(1986)Ceperley, and Alder]{QMC}
Ceperley,~D.; Alder,~B. Quantum {Monte Carlo}. \emph{Science} \textbf{1986},
  \emph{231}, 555--560\relax
\mciteBstWouldAddEndPuncttrue
\mciteSetBstMidEndSepPunct{\mcitedefaultmidpunct}
{\mcitedefaultendpunct}{\mcitedefaultseppunct}\relax
\EndOfBibitem
\bibitem[White(1992)]{DMRG}
White,~S.~R. Density matrix formulation for quantum renormalization groups.
  \emph{Phys. Rev. Lett.} \textbf{1992}, \emph{69}, 2863--2866\relax
\mciteBstWouldAddEndPuncttrue
\mciteSetBstMidEndSepPunct{\mcitedefaultmidpunct}
{\mcitedefaultendpunct}{\mcitedefaultseppunct}\relax
\EndOfBibitem
\bibitem[Kent and Kotliar(2018)Kent, and Kotliar]{DMFT}
Kent,~P. R.~C.; Kotliar,~G. Toward a predictive theory of correlated materials.
  \emph{Science} \textbf{2018}, \emph{361}, 348--354\relax
\mciteBstWouldAddEndPuncttrue
\mciteSetBstMidEndSepPunct{\mcitedefaultmidpunct}
{\mcitedefaultendpunct}{\mcitedefaultseppunct}\relax
\EndOfBibitem
\bibitem[Motta \latin{et~al.}(2017)Motta, Ceperley, Chan, Gomez, Gull, Guo,
  Jim\'enez-Hoyos, Lan, Li, Ma, Millis, Prokof'ev, Ray, Scuseria, Sorella,
  Stoudenmire, Sun, Tupitsyn, White, Zgid, and Zhang]{Motta2017}
Motta,~M. \latin{et~al.}  Towards the solution of the many-electron problem in
  real materials: Equation of state of the hydrogen chain with state-of-the-art
  many-body methods. \emph{Phys. Rev. X} \textbf{2017}, \emph{7}, 031059\relax
\mciteBstWouldAddEndPuncttrue
\mciteSetBstMidEndSepPunct{\mcitedefaultmidpunct}
{\mcitedefaultendpunct}{\mcitedefaultseppunct}\relax
\EndOfBibitem
\bibitem[Williams \latin{et~al.}(2020)Williams, Yao, Li, Chen, Shi, Motta, Niu,
  Ray, Guo, Anderson, Authors, and Authors]{Zgid}
Williams,~K.~T.; Yao,~Y.; Li,~J.; Chen,~L.; Shi,~H.; Motta,~M.; Niu,~C.;
  Ray,~U.; Guo,~S.; Anderson,~R.~J.; Authors,~M.; Authors,~M. Direct comparison
  of many-body methods for realistic electronic {Hamiltonians}. \emph{Phys.
  Rev. X} \textbf{2020}, \emph{10}, 011041\relax
\mciteBstWouldAddEndPuncttrue
\mciteSetBstMidEndSepPunct{\mcitedefaultmidpunct}
{\mcitedefaultendpunct}{\mcitedefaultseppunct}\relax
\EndOfBibitem
\bibitem[Burke(2012)]{Burke}
Burke,~K. Perspective on density functional theory. \emph{J. Chem. Phys.}
  \textbf{2012}, \emph{136}, 150901\relax
\mciteBstWouldAddEndPuncttrue
\mciteSetBstMidEndSepPunct{\mcitedefaultmidpunct}
{\mcitedefaultendpunct}{\mcitedefaultseppunct}\relax
\EndOfBibitem
\bibitem[Gaitan and Nori(2009)Gaitan, and Nori]{Nori}
Gaitan,~F.; Nori,~F. Density functional theory and quantum computation.
  \emph{Phys. Rev. B} \textbf{2009}, \emph{79}, 205117\relax
\mciteBstWouldAddEndPuncttrue
\mciteSetBstMidEndSepPunct{\mcitedefaultmidpunct}
{\mcitedefaultendpunct}{\mcitedefaultseppunct}\relax
\EndOfBibitem
\bibitem[Tempel and Aspuru-Guzik(2012)Tempel, and Aspuru-Guzik]{Guzik}
Tempel,~D.~G.; Aspuru-Guzik,~A. Quantum computing without wavefunctions:
  Time-dependent density functional theory for universal quantum computation.
  \emph{Sci. Rep.} \textbf{2012}, \emph{2}, 391\relax
\mciteBstWouldAddEndPuncttrue
\mciteSetBstMidEndSepPunct{\mcitedefaultmidpunct}
{\mcitedefaultendpunct}{\mcitedefaultseppunct}\relax
\EndOfBibitem
\bibitem[Hidary(2019)]{Hidari}
Hidary,~J.~D. \emph{Quantum Computing: an Applied Approach}; Springer,
  2019\relax
\mciteBstWouldAddEndPuncttrue
\mciteSetBstMidEndSepPunct{\mcitedefaultmidpunct}
{\mcitedefaultendpunct}{\mcitedefaultseppunct}\relax
\EndOfBibitem
\bibitem[Boykin \latin{et~al.}(2000)Boykin, Mor, Pulver, Roychowdhury, and
  Vatan]{clifft}
Boykin,~P.~O.; Mor,~T.; Pulver,~M.; Roychowdhury,~V.; Vatan,~F. A new universal
  and fault-tolerant quantum basis. \emph{Inf. Process. Lett.} \textbf{2000},
  \emph{75}, 101--107\relax
\mciteBstWouldAddEndPuncttrue
\mciteSetBstMidEndSepPunct{\mcitedefaultmidpunct}
{\mcitedefaultendpunct}{\mcitedefaultseppunct}\relax
\EndOfBibitem
\bibitem[Boixo \latin{et~al.}(2018)Boixo, Isakov, Smelyanskiy, Babbush, Ding,
  Jiang, Bremner, Martinis, and Neven]{Boixo}
Boixo,~S.; Isakov,~S.~V.; Smelyanskiy,~V.~N.; Babbush,~R.; Ding,~N.; Jiang,~Z.;
  Bremner,~M.~J.; Martinis,~J.~M.; Neven,~H. Characterizing quantum supremacy
  in near-term devices. \emph{Nat. Phys.} \textbf{2018}, \emph{14}, 595--600,
  {The} QCs used in this work were taken from the first author's
  \href{https://github.com/sboixo/GRCS/tree/master/inst/rectangular/cz_v2}{Github
  account}, last accessed in January 2021\relax
\mciteBstWouldAddEndPuncttrue
\mciteSetBstMidEndSepPunct{\mcitedefaultmidpunct}
{\mcitedefaultendpunct}{\mcitedefaultseppunct}\relax
\EndOfBibitem
\bibitem[Peruzzo \latin{et~al.}(2014)Peruzzo, McClean, Shadbolt, Yung, Zhou,
  Love, Aspuru-Guzik, and O’brien]{VQE}
Peruzzo,~A.; McClean,~J.; Shadbolt,~P.; Yung,~M.-H.; Zhou,~X.-Q.; Love,~P.~J.;
  Aspuru-Guzik,~A.; O’brien,~J.~L. A variational eigenvalue solver on a
  photonic quantum processor. \emph{Nat. Commun.} \textbf{2014}, \emph{5},
  1--7\relax
\mciteBstWouldAddEndPuncttrue
\mciteSetBstMidEndSepPunct{\mcitedefaultmidpunct}
{\mcitedefaultendpunct}{\mcitedefaultseppunct}\relax
\EndOfBibitem
\bibitem[Fedorov \latin{et~al.}(2022)Fedorov, Peng, Govind, and Alexeev]{VQE-2}
Fedorov,~D.~A.; Peng,~B.; Govind,~N.; Alexeev,~Y. {VQE} method: A short survey
  and recent developments. \emph{Mater. Theory} \textbf{2022}, \emph{6},
  1--21\relax
\mciteBstWouldAddEndPuncttrue
\mciteSetBstMidEndSepPunct{\mcitedefaultmidpunct}
{\mcitedefaultendpunct}{\mcitedefaultseppunct}\relax
\EndOfBibitem
\bibitem[Mazziotti(2012)]{mazziotti-1}
Mazziotti,~D.~A. Two-electron reduced density matrix as the basic variable in
  many-electron quantum chemistry and physics. \emph{Chem. Rev.} \textbf{2012},
  \emph{112}, 244--262\relax
\mciteBstWouldAddEndPuncttrue
\mciteSetBstMidEndSepPunct{\mcitedefaultmidpunct}
{\mcitedefaultendpunct}{\mcitedefaultseppunct}\relax
\EndOfBibitem
\bibitem[Mazziotti(2006)]{Mazziotti-2}
Mazziotti,~D.~A. Anti-Hermitian contracted Schr\"odinger equation: Direct
  determination of the two-electron reduced density matrices of many-electron
  molecules. \emph{Phys. Rev. Lett.} \textbf{2006}, \emph{97}, 143002\relax
\mciteBstWouldAddEndPuncttrue
\mciteSetBstMidEndSepPunct{\mcitedefaultmidpunct}
{\mcitedefaultendpunct}{\mcitedefaultseppunct}\relax
\EndOfBibitem
\bibitem[Snyder \latin{et~al.}(2012)Snyder, Rupp, Hansen, M\"uller, and
  Burke]{ML-DFT}
Snyder,~J.~C.; Rupp,~M.; Hansen,~K.; M\"uller,~K.-R.; Burke,~K. Finding Density
  Functionals with Machine Learning. \emph{Phys. Rev. Lett.} \textbf{2012},
  \emph{108}, 253002\relax
\mciteBstWouldAddEndPuncttrue
\mciteSetBstMidEndSepPunct{\mcitedefaultmidpunct}
{\mcitedefaultendpunct}{\mcitedefaultseppunct}\relax
\EndOfBibitem
\bibitem[Pederson \latin{et~al.}(2022)Pederson, Kalita, and
  Burke]{Review-ML-DFT}
Pederson,~R.; Kalita,~B.; Burke,~K. Machine learning and density functional
  theory. \emph{Nat. Rev. Phys.} \textbf{2022}, \emph{4}, 357\relax
\mciteBstWouldAddEndPuncttrue
\mciteSetBstMidEndSepPunct{\mcitedefaultmidpunct}
{\mcitedefaultendpunct}{\mcitedefaultseppunct}\relax
\EndOfBibitem
\bibitem[Kay(2020)]{tikz}
Kay,~A. Tutorial on the {Quantikz} Package. \emph{arXiv 1809.03842}
  \textbf{2020}, \relax
\mciteBstWouldAddEndPunctfalse
\mciteSetBstMidEndSepPunct{\mcitedefaultmidpunct}
{}{\mcitedefaultseppunct}\relax
\EndOfBibitem
\end{mcitethebibliography}
\providecommand{\latin}[1]{#1}
\makeatletter
\providecommand{\doi}
  {\begingroup\let\do\@makeother\dospecials
  \catcode`\{=1 \catcode`\}=2 \doi@aux}
\providecommand{\doi@aux}[1]{\endgroup\texttt{#1}}
\makeatother
\providecommand*\mcitethebibliography{\thebibliography}
\csname @ifundefined\endcsname{endmcitethebibliography}
  {\let\endmcitethebibliography\endthebibliography}{}

\end{document}